\documentclass[runningheads]{llncs}
\usepackage[T1]{fontenc}

\usepackage{graphicx}
\usepackage{diagbox}
\usepackage{microtype}
\usepackage{siunitx}
\usepackage{subcaption}
\usepackage{float} 
\begin{document}
\title{Enhancing Speaker Verification with Whispered Speech via Post-Processing}
%
%
\author{Magdalena Gołębiowska\inst{}\orcidID{0000-0001-7507-6737} \and
Piotr~Syga\inst{}\orcidID{0000-0002-0266-5802}}
\authorrunning{M. Gołębiowska \and P. Syga}
%
\institute{Department of Artificial Intelligence,\\Wroclaw University of Science and Technology,\\ Wybrzeze Wyspianskiego 27, Wroclaw, 50-370, Poland\\
\email{\{magdalena.golebiowska,piotr.syga\}@pwr.edu.pl}}

\pagestyle{empty}
\begin{center}
    \textbf{\Large IEEE Copyright Notice}
\end{center}

\noindent \copyright 2026 IEEE. Personal use of this material is permitted. Permission from IEEE must be obtained for all other uses, in any current or future media, including reprinting/republishing this material for advertising or promotional purposes, creating new collective works, for resale or redistribution to servers or lists, or reuse of any copyrighted component of this work in other works.
\vspace{2cm}

\noindent This work has been accepted for publication in 18th Asian Conference on Intelligent Information and Database Systems. The final published version will be available via IEEE Xplore.

\newpage
\maketitle              
\begin{abstract} 
Speaker verification is a task of confirming an individual's identity through the analysis of their voice. Whispered speech differs from phonated speech in acoustic characteristics, which degrades the performance of speaker verification systems in real-life scenarios, including avoiding fully phonated speech to protect privacy, disrupt others, or the lack of full vocalization may be dictated by a disease. In this paper we propose a model with a training recipe to obtain more robust representations against whispered speech hindrances. The proposed system employs an encoder–decoder structure built atop a fine-tuned speaker verification backbone, optimized jointly using cosine similarity–based classification and triplet loss. We gain relative improvement of 22.26\% compared to the baseline (baseline 6.77\% vs ours 5.27\%) in normal vs whispered speech trials, achieving AUC of 98.16\%. In tests comparing whispered to whispered, our model attains an EER of 1.88\% with AUC equal to 99.73\%, which represents a 15\% relative enhancement over the prior leading ReDimNet-B2. We also offer a summary of the most popular and state-of-the-art speaker verification models in terms of their performance with whispered speech. Additionally, we evaluate how these models perform under noisy audios, obtaining that generally the same relative level of noise degrades the performance of speaker verification more significantly on whispered speech than on normal speech.
\end{abstract}
\section{Introduction}

This study examines speaker verification (SV) with whispered speech. Speaker verification is a task of confirming an individual's identity through the analysis of their speech to determine whether to accept or deny the claimed identity of the speaker~\cite{JUANG2003485}. The testing sample from the speaker is compared to another previously obtained sample, usually recorded by a user in a neutral state, both emotionally and vocally, i.e., in fully phonated speech. 

Acoustic characteristics of whispered speech are different from those of normal speech due to the absence of vocal cord vibrations \cite{JOVICIC2008263}, including an upward shift of formant frequencies of vowels, lower energy on voiced consonants at low frequencies, and greater spectral flatness \cite{ITO2005139}. Naturally, these varying vocal efforts cause a mismatch between a previously obtained speaker sample in a neutral environment and a new testing sample. These are problematic for speaker verification systems that have not been designed and evaluated for whispered speech \cite{Sarria-Paja_Falk_2015}. Considering whispered speech data in speaker verification is motivated by real-world use cases. A user might whisper contents to protect privacy, to avoid disrupting others, or the change is forced by disease or vocal cords are removed after an operation \cite{JOVICIC2008263}.

Previous studies on whispered speech speaker verification primarily relied on legacy architectures that predate recent advances in deep embedding systems.

First approaches were based on including whispered speech in the training data for a GMM-UBM model \cite{6639062}, which improved the performance of speaker verification with whispered speech. Additionally, the work proved that using speaking-style and gender dependent-models, and adding AM-FM signal representation-based features also improved verification reliability. 

The study expanded into testing diverse approaches, including frequency warping, sub-band analysis, alternate feature representations, and feature combination. The findings revealed that these modifications were ineffective in improving whispered data, highlighting the necessity for representations tailored to whispered speech. The most successful model in the study achieved an EER of 22.77\% on the CHAINS (Characterizing Individual Speakers) dataset~\cite{chains} under normal versus whispered speech trials.

Such features were proposed by fusing information from
spectral, modulation spectral and bottleneck features computed via deep neural networks at the feature- and score-levels \cite{SARRIAPAJA2017437}. The experimental findings indicated that relative enhancements could reach up to 79\% for neutral speech and 60\% for whispered speech on wTIMIT and TIMIT when compared to a baseline system trained using i-vectors derived from mel frequency cepstral coefficients. These findings were further improved by introducing three distinct features and implementing a score fusion method, which relied on systems trained on these three feature sets \cite{SARRIAPAJA201878}, resulting in 66\% and 63\% relative improvement on combined CHAINS, TIMIT, and wTIMIT databases for whispered and normal speech, respectively.

Another approach to choosing features robust to whispered speech was based on formant and formant gap (FoG) features thought to be more invariant to speech modes than MFCCs and AAMFs \cite{8682571}. The system was trained on i-vectors and showed a 3.79\% absolute value improvement in EER on CHAINS, TIMIT, and wTIMIT compared to the baseline AAMF in the mismatched trials; however, under normal conditions, the performance slightly degrades.

The important notion is that all mentioned research until now has been conducted on systems that required a development phase, i.e., they have seen the speaker data during training, which entails no generalizability to unseen speakers, which is not the correct approach nowadays.

The most recent studies explored the x-vector/PLDA/SPLICE system \cite{PRIETO2022103536}, which scored 17.8\% EER in neutral-whispered conditions on CHAINS, resulting in a 0.5\% relative improvement compared to the baseline. 

Semi-supervised learning models were also explored in the context of SV with whispered speech. A system based on wav2vec 2.0 with augmentation based on whisper synthesis using Praat software and domain score normalization with whisper detection achieved 2.29\% EER on CHAINS in whisper vs whisper trials and 46.31\% relative improvement on the MSP-AVW in normal vs whisper conditions \cite{10977907}.

In this article, we propose a new model with a training framework for  more robust embeddings against whispered speech. We provide a review of robustness against whispered speech of recent and state-of-the-art speaker verification encoders. Additionally, we check how noise affects the baseline models' performance. Our contributions are as follows:
\begin{enumerate}
    \item We propose a new model and training for speaker verification which improves verification performance with whispered speech.
    \item To our knowledge, this is the first study to evaluate speaker verification with whispered speech on state-of-the-art speaker verification systems, including ECAPA2, ECAPA-TDNN, and different versions of ReDimNet.
    \item We evaluate how state-of-the-art speaker verification models act under noisy whispered speech.
\end{enumerate}

\section{System Architecture and Training}
In this section, we present details of our proposed architecture and training recipe for speaker verification with whispered speech.

To improve speaker verification with whispered speech, we introduce a few fully-connected layers with ReLU that create an encoder-decoder architecture with bootleneck on top of the pre-trained speaker verification model chosen to be ReDimNet-B6 \cite{yakovlev24_interspeech}.  The encoder-decoder module is shallow by design with limited capacity, consisting of four fully-connected layers. The choice is dictated by the already employed strong speaker recognition model (ReDimNet-B6); thus, its objective is to only compensate for the systematic phonation mismatch between whispered and phonated speech.  Preliminary experiments with deeper or wider architectures resulted in either no additional gains or degraded performance, indicating that higher model capacity tends to over-adapt to the training speakers. The bottleneck dimensionality was chosen to enforce a compact latent representation that preserves phonation-related variability while filtering out nuisance factors. This design encourages the network to act as a residual correction rather than a complete transformation of the embedding. 

\begin{figure}
\includegraphics[width=\textwidth]{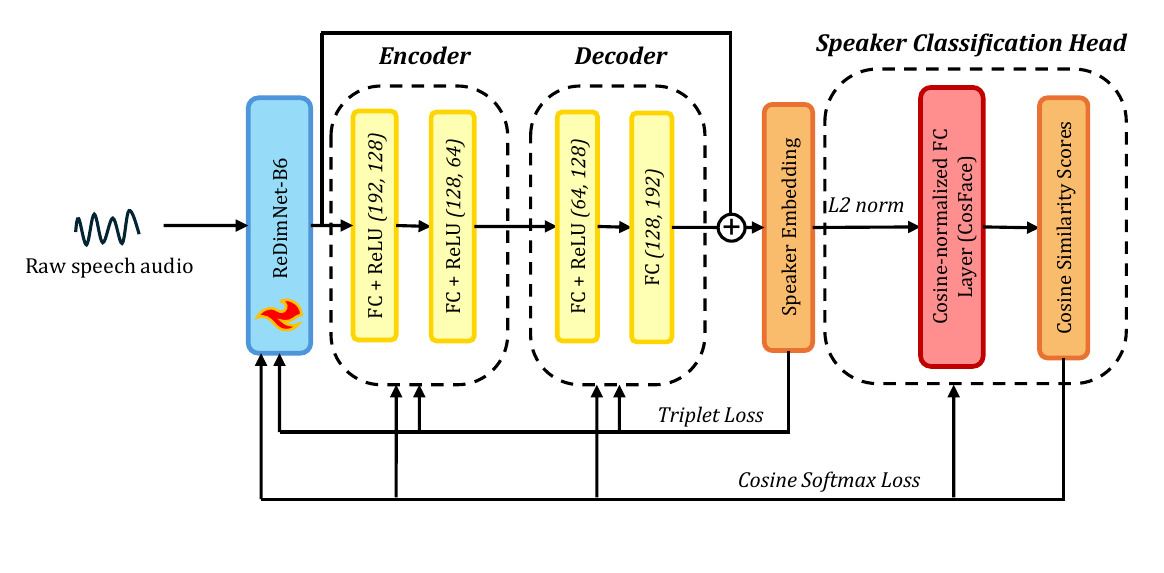}
\caption{Proposed architecture during training phase. The proposed architecture employs an encoder and decoder layers which map whispered representations into phonated ones and a speaker classification head with cosine-normalized linear projection scaled by a constant factor and optimized using cross-entropy loss and triplet loss.} \label{fig:train-arch}
\end{figure}

To preserve speaker identity, we add a speaker classification head that consists of a cosine-normalized fully-connected layer and outputs cosine similarity scores as seen in NormFace \cite{Wang2017NormFaceLH}. Fig.~\ref{fig:train-arch} presents the model training. The audio embeddings processed by the encoder-decoder are compared using triplet loss in triplets (phonated sentence, whispered sentence, random sentence chosen from another speaker) \(L_{trip}\). We also add a residual connection between encoder input and encoder output. The objective for the model is to learn to convert whispered representations into phonated ones while conserving speaker identity. To further keep speaker discerning capabilities, we add cosine softmax loss \(L_{ce}\) with similarities coming from the speaker classification head. The losses are combined with formula:

\begin{equation}
     L = L_{trip} + \gamma  L_{ce}~,
\end{equation}

\noindent where \(\gamma=10^{-4}\). The value was selected to scale \(L_{ce}\) accordingly to \(L_{trip}\) to influence model weights less than the triplet loss. The pre-trained speaker verification model is fine tuned with layers gradually unfrozen every 5 epochs.

\begin{figure}
\includegraphics[width=\textwidth]{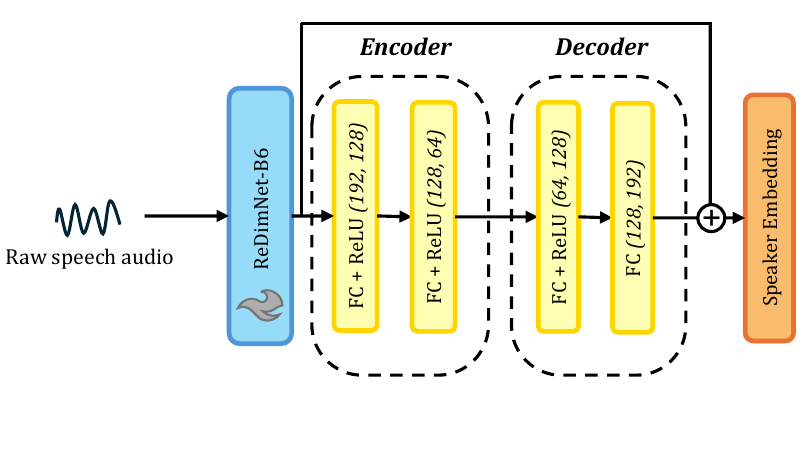}
\caption{Proposed architecture during evaluation phase. Speaker classification head is detached and the audio passes through fine-tuned ReDimNet whose representations are further processed by encoder and decoder to convert whispered representations into phonated ones.} \label{fig:test-arch}
\end{figure}

After training, we remove the speaker classification head, as shown in Fig.~\ref{fig:test-arch}, leaving only the encoder and decoder that produce embeddings used for speaker verification. At this point all weights are frozen. 

\section{Experimental Setup}

\subsection{Dataset}
We evaluate the systems using the CHAINS (Characterizing Individual Speakers) database \cite{chains} released in 2008. This dataset features 36 English speakers (18 male, 18 female) who read fables and selected sentences employing various speaking styles, including neutral and whispered. Data gathering occurred under laboratory conditions. We utilized subsets of the dataset described as solo and whisper readings of fables and sentences (text ids f01-f04, sentences s01-s33), resulting in a total of 5,860 samples (158-198 samples per speaker), balanced between neutral and whispered speech as we excluded audios without a pair from both speaking styles.  Measured average root-mean-square energy (RMSE) value for normal speech \(\mu_{norm} = 0.046, \ \sigma_{norm} = 0.014\) and for whispered speech \(\mu_{whsp} = 0.0138, \ \sigma_{whsp} = 0.0032\). As expected, energy of whispered speech is lower indicating lower perceived loudness.

We split the dataset by speakers, using 70\% for training and 30\% for testing. We keep the split in the speaker verification tests.

For noise robustness experiments, we use MUSAN (Music, Speech, and Noise Corpus) \cite{musan2015}. It consists of  approximately 109 hours of audio in the US Public Domain or under a Creative Commons license. The audios are categorized in music, speech and noise. We use the noise part of the corpus with the duration of 6 hours. Noises consist of technical sounds, including DTMF tones, dial tones, and fax machine sounds, as well as environmental sounds like idling cars, thunder, wind, footsteps, rustling paper, rain, various animal noises and crowd noises.

\subsection{Environment and Training}
We reran our experiments a total of 10 times to assess the model's performance and consistency, then computed the average of the outcomes. Each experiment was run on a single NVIDIA Hopper (H100) GPU with 955 GB RAM (not used to full capacity). Code base with details for this paper is available in a Github repository\footnote{Code base is available at https://github.com/mgraves236/sv-whispred-speech}.

Audio samples underwent preprocessing to meet the front-end requirements, which included converting them to mono-channel and resampling to \SI{16}{\kilo\hertz}.

The system was trained with an Adam optimizer (learning rate of $10^{-4}$ for encoder-decoder and speaker classification head, and $10^{-5}$ for fine-tuning ReDimNet-B6, weight decay of $10^{-4}$) for 100 epochs, batch size 128.  We added dropout with value 0.3 to encoder-decoder layers.

In order to evaluate, we randomly select pairs for each dataset sample. For each instance, we randomly select one positive sample, originating from the same speaker, and one negative sample, which comes from a different speaker. The trial type being assessed determines the vocal style of these samples.

We evaluate our framework using the Equal Error Rate (EER), a standard metric in speaker verification
tasks. EER corresponds to the point where the misclassification rates of positive
and negative samples are equal (i.e., FPR=FNR), and lower values indicate better overall performance.

\section{Results}

Tab.~\ref{tab:results-main} presents the EER scores for the testing subset of the dataset across various trials. These trials consider different speaking conditions between enrollment and testing, such as the common situation where a user enrolls with normal speech and later attempts to authenticate using whispered speech.

We include the most popular and state-of-the-art speaker verification systems, i.e. x-vector \cite{8461375} as shared by SpeechBrain Toolkit \cite{speechbrain}, ECAPA-TDNN \cite{ecapa-tdnn} also from SpeechBrain \cite{speechbrain}, ECAPA2 \cite{ecapa2} shared by the original authors, ReDimNet-B0, ReDimNet-B2, and ReDimNet-B6 \cite{yakovlev24_interspeech} pretrained by the original authors.

\begin{table}[t]
\centering
\caption{The comparison of EER scores on the testing dataset of baselines and our model across trials of normal vs whispered speech (Norm vs Whsp), normal vs normal speech (Norm vs Norm), whispered vs whispered speech (Whsp vs Whsp), both whispered and normal vs whispered and normal speech (All vs All). The table lists mean and standard deviation from 10 independent reruns. The best scores are in bold. }
\vspace{0.5em}
\begin{tabular}{|l|c|c|c|c|}
\hline
 \diagbox[width=2.2cm,height=0.7cm]{\textbf{Model}}{\textbf{Trial}} &  \textbf{Norm vs Whsp} & \textbf{Norm vs Norm} &  \textbf{Whsp vs Whsp} & \textbf{All vs All}\\
\hline

x-vector 
&  \begin{tabular}{@{}c@{}}26.56\% \\$\pm$ 1.44 \% \end{tabular} 
&  \begin{tabular}{@{}c@{}}3.32\% \\$\pm$ 0.59 \% \end{tabular}  
& \begin{tabular}{@{}c@{}}8.48\% \\$\pm$ 0.94 \% \end{tabular}
&\begin{tabular}{@{}c@{}}29.93\% \\$\pm$ 1.44 \% \end{tabular} 	 \rule{0pt}{5ex}\\ 

ECAPA-TDNN 
& \begin{tabular}{@{}c@{}}10.49\% \\$\pm$ 1.47 \% \end{tabular}   
& \begin{tabular}{@{}c@{}}0.41 \% \\$\pm$ 0.18 \% \end{tabular} 
& \begin{tabular}{@{}c@{}}2.77\% \\$\pm$ 0.51 \% \end{tabular}  
& \begin{tabular}{@{}c@{}}13.72\% \\$\pm$ 1.26 \% \end{tabular}	\rule{0pt}{5ex}\\

ECAPA2 
& \begin{tabular}{@{}c@{}}8.28\% \\$\pm$ 0.76 \% \end{tabular}   
& \begin{tabular}{@{}c@{}}0.21 \% \\$\pm$ 0.24 \% \end{tabular} 
& \begin{tabular}{@{}c@{}}2.48\% \\$\pm$ 0.53 \% \end{tabular}  
& \begin{tabular}{@{}c@{}}10.95\% \\$\pm$ 0.75 \% \end{tabular}	\rule{0pt}{5ex}\\

ReDimNet-B0 
& \begin{tabular}{@{}c@{}}13.02\% \\$\pm$ 1.56 \% \end{tabular}   
& \begin{tabular}{@{}c@{}}0.49 \% \\$\pm$ 0.31 \% \end{tabular} 
& \begin{tabular}{@{}c@{}}2.38\% \\$\pm$ 0.40 \% \end{tabular}  
& \begin{tabular}{@{}c@{}}19.57\% \\$\pm$ 0.96 \% \end{tabular}	\rule{0pt}{5ex}\\

ReDimNet-B2 
& \begin{tabular}{@{}c@{}}8.80\% \\$\pm$ 1.22 \% \end{tabular}   
& \begin{tabular}{@{}c@{}}0.23 \% \\$\pm$ 0.16 \% \end{tabular} 
& \begin{tabular}{@{}c@{}}2.20\% \\$\pm$ 0.38 \% \end{tabular}  
& \begin{tabular}{@{}c@{}}12.75\% \\$\pm$ 0.88 \% \end{tabular}	\rule{0pt}{5ex}\\

ReDimNet-B6 
& \begin{tabular}{@{}c@{}}6.77\% \\$\pm$ 1.51 \% \end{tabular}   
& \begin{tabular}{@{}c@{}} \textbf{0.12\%} \\$\pm$ 0.16 \% \end{tabular} 
& \begin{tabular}{@{}c@{}}2.30\% \\$\pm$ 0.43 \% \end{tabular}  
& \begin{tabular}{@{}c@{}} \textbf{7.76\%} \\$\pm$ 0.78 \% \end{tabular}	\rule{0pt}{5ex}\\

\textbf{Ours} 

& \begin{tabular}{@{}c@{}} \textbf{5.27\%} \\$\pm$ 0.82 \% \end{tabular}   
& \begin{tabular}{@{}c@{}}0.28 \% \\$\pm$ 0.20 \% \end{tabular} 
& \begin{tabular}{@{}c@{}} \textbf{1.88\%} \\$\pm$ 0.28 \% \end{tabular}  
& \begin{tabular}{@{}c@{}}8.40\% \\$\pm$ 0.90 \% \end{tabular}	\rule{0pt}{5ex}\\

\hline 
\end{tabular}
\label{tab:results-main}
\end{table}

Among the models evaluated, x-vector showed the lowest robustness, with an EER of 29.93\% in the All vs All trial. ReDimNet-B2 achieved an EER of 12.75\%, which was similar to ECAPA-TDNN's 13.72\%. Its successor, ECAPA2, performed marginally better, with an EER of 10.95\%. The top performer was ReDimNet-B6, with an EER of 7.76\%. Our model achieved second best result of EER 8.40\%. The likely reason for this drop in performance is that the training dataset was not as large as the original datasets used during pre-training.

In Norm vs Norm conditions this trend also applies. X-vector as the oldest system, performs the worst with EER equal to 3.32\%. ECAPA-TDNN and ReDimNet-B0 were comparable in this conditions achieving EER of 0.41\% and 0.49\% respectively. Similar results were achieved for ECAPA2, ReDimNet-B2, and by our model with EER of 0.21\%, 0.23\%, and 0.28\% accordingly.

In Whsp vs Whsp trials, the EERs are lower compared to Norm vs Whsp tests due to the absence of mixed conditions, yet they remain higher than in Norm vs Norm trials. Again, the worst performance of 8.48\% was noted in x-vector. Remaining models, ECAPA-TDNN, ECAPA2, ReDimNet-B0, ReDimNet-B2, and ReDimNet-B6 showed similar EER of 2.77\%, 2.48\%, 2.38\%, 2.20\%, and 2.30\% respectively. Interestingly, ReDimNet-B6, which surpassed other ReDimNets in other trials, was 0.10 percentage points worse than its smaller version ReDimNet-B2. Our model achieved the best result in this category with EER equal to 1.88\%, i.e, 15\% relative improvement over the previous best ReDimNet-B2.

Among all evaluation conditions, the Normal vs Whispered trial is the most challenging and realistic verification scenario. Our fine-tuned ReDimNet-B6 reduces EER from 6.77\% to 5.27\%, a 22.16\% relative improvement over the unadapted model, indicating that the proposed fine-tuning approach substantially improves robustness to cross-phonation variability.

Although the proposed model slightly underperforms in the Norm vs Norm and All vs All conditions, this is consistent with its design objective: enhancing robustness to phonation mismatch. The trade-off is minimal and justified by the significant improvement in Normal vs Whispered performance.

\begin{figure}[H]
\centering
\subcaptionbox{ROC for Normal vs Whisper trial}{\includegraphics[width=0.5\textwidth]{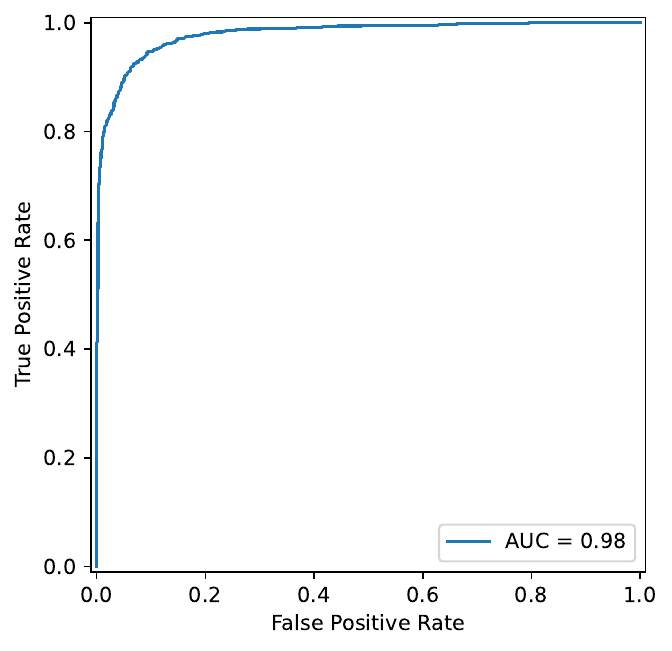}}%
\hfill
\subcaptionbox{ROC for Normal vs Normal trial}{\includegraphics[width=0.5\textwidth]{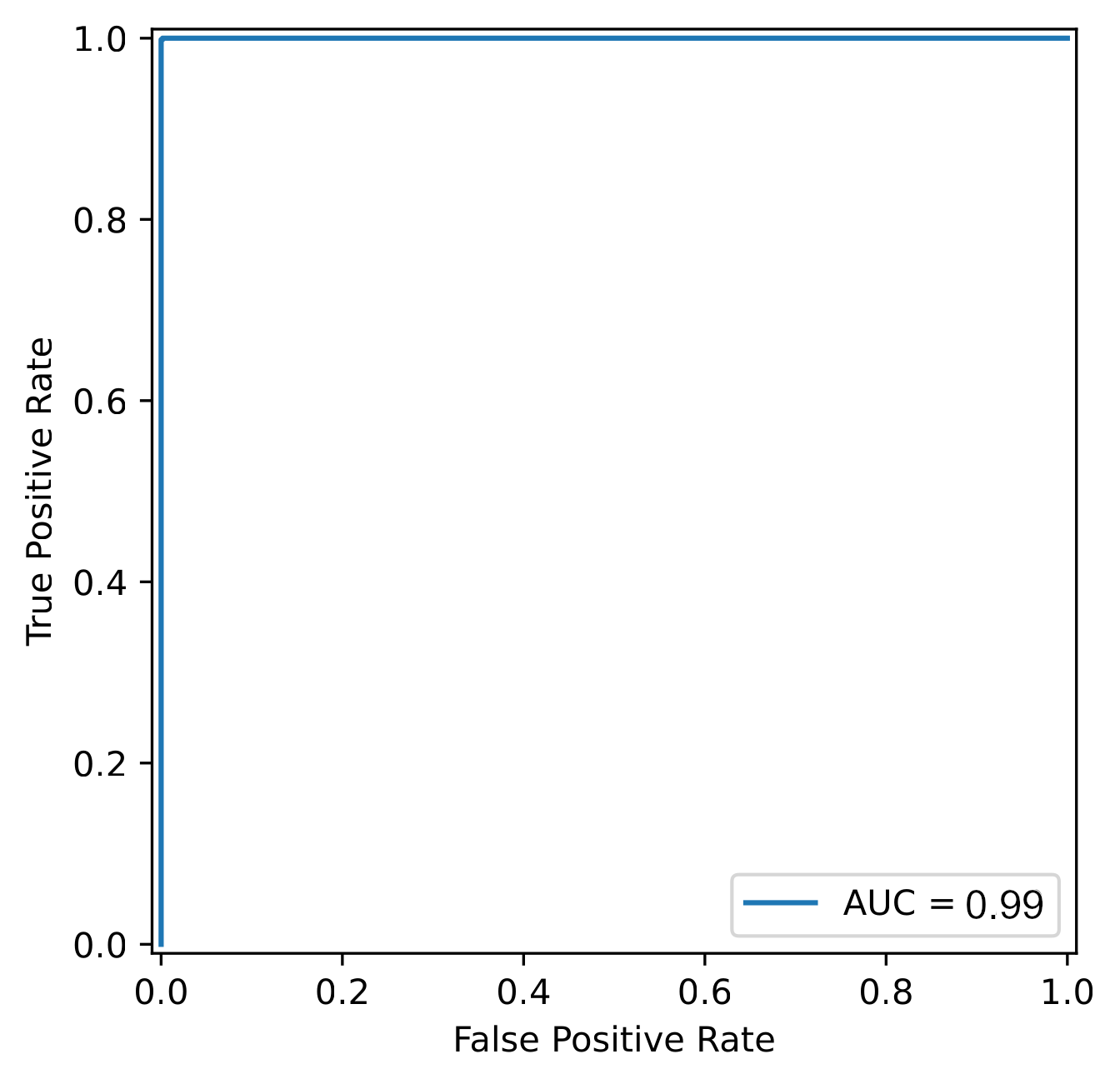}}%
\hfill
\subcaptionbox{ROC for Whisper vs Whisper trial}{\includegraphics[width=0.5\textwidth]{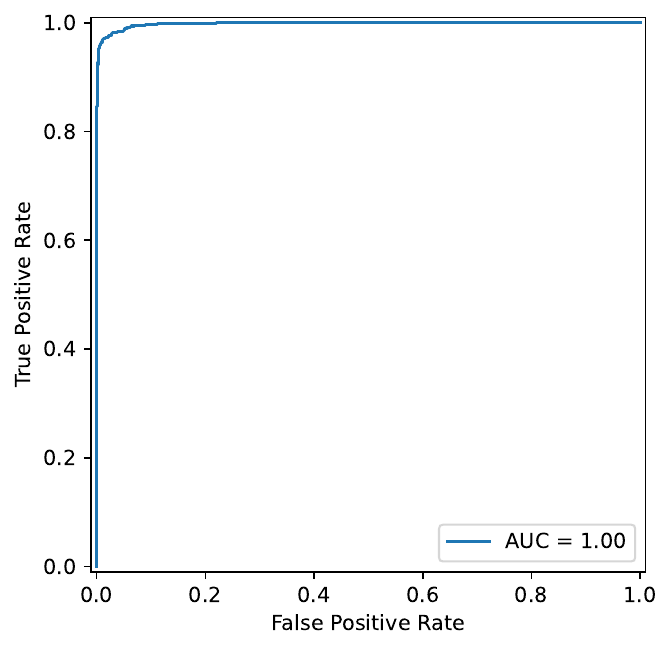}}%
\hfill
\subcaptionbox{ROC for All vs All trial}{\includegraphics[width=0.5\textwidth]{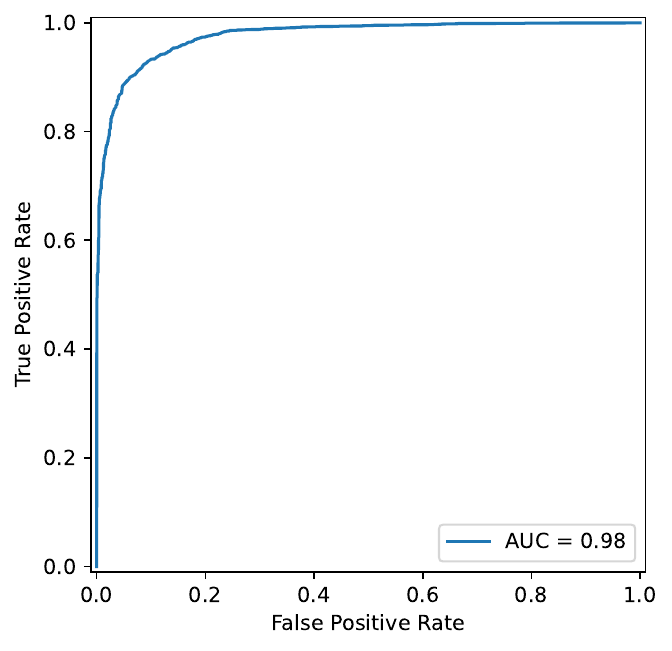}}%
\caption{ROC curves illustrating performance of our model from one run (one seed) throughout all trials. Area under the curve is displayed in the bottom right of each figure. AUC close to 1 shows that our model acts almost perfectly.}
\label{fig:roc}
\end{figure}

To further assess the efficiency of our model, we inspected receiver operating characteristic curves across all trails (Fig.~\ref{fig:roc}) and area under curve values (Tab.~\ref{tab:results-auc}). For Normal vs Whisper tests, the curve is not ideal, however still very steep, reaching high True Positive Rate at low False Positive Rate, thus the model is able to identify most positives early while keeping false alarms at a very low level. Averaged AUC is equal to \(98.16\%\) which is very close to a perfect score. For Normal vs Normal, the ROC is almost ideal which is supported by the averaged AUC equal to \(99.99\%\), meaning the model raises very few false alarms. In Whisper versus Whisper trials, the performance slightly declines, as evident in the form of the ROC curve and the averaged AUC value of 99.73\%. For the All vs All AUC, the lowest value is 97.72\%, as reflected in the shape of the ROC curve. We have to take into the account that the mono-style trials i.e. Normal vs Normal and Whisper vs Whisper, due to low data availability cover 3,024 samples, which is a relatively mediocre amount.

\begin{table}[h]
\centering
\caption{The comparison of AUC scores on the testing dataset across trials of normal vs whispered speech (Norm vs Whsp), normal vs normal speech (Norm vs Norm), whispered vs whispered speech (Whsp vs Whsp), both whispered and normal vs whispered and normal speech (All vs All). The table lists mean and standard deviation from 10 independent runs.}
\vspace{0.5em}
\hfill
\begin{tabular}{|l|c|c|c|c|}
\hline
 \diagbox[width=2.2cm,height=0.7cm]{\textbf{Model}}{\textbf{Trial}} &  \textbf{Norm vs Whsp} & \textbf{Norm vs Norm} &  \textbf{Whsp vs Whsp} & \textbf{All vs All}\\
\hline
\textbf{Ours} 
& \begin{tabular}{@{}c@{}} \textbf{98.16\%} \\$\pm$ 0.48 \% \end{tabular}   
& \begin{tabular}{@{}c@{}}\textbf{100.0\%} \\$\pm$ 0.00 \% \end{tabular} 
& \begin{tabular}{@{}c@{}} \textbf{99.73\%} \\$\pm$ 0.08 \% \end{tabular}  
& \begin{tabular}{@{}c@{}}\textbf{97.72\%} \\$\pm$ 0.36 \% \end{tabular}\rule{0pt}{5ex}\\

\hline 
\end{tabular}
\label{tab:results-auc}
\end{table}

\subsection{Ablation Study}
Additionally, we conducted a few experiments to investigate which part of the architecture provided the majority of gain.
Results are listed in Tab.~\ref{tab:results-ablation}.

The first experiment involved changing the ReDimNet-B6 front end to ECAPA-TDNN. The experiments were run with learning rate equal to $10^{-4}$, as it gave better performance than with previous learning rate of $10^{-5}$. We also kept the idea of unfreezing the layers gradually every 5 epochs. Still, using ECAPA-TDNN as an embeddings extractor, gave worse results than ReDimNet-B6 (14.20\% vs 5.27\% for All vs All). Interestingly, the results are worse than the baseline (14.20\% vs 13.72\% for All vs All), suggesting that more parameter-tuning would be needed. 

Next, we unfroze only the last two blocks of ReDimNet-B6, allowing only the last layers to adapt to the new domain. This way, the achieved results were overall worse than our proposed model (9.19\% vs 8.40\% in All vs All) but the model was slightly better in Norm vs Norm trial (0.27\% vs 0.28\%), showing that unfreezing more layers degrades performance for normal speech speaker verification. This is related to catastrophic forgetting of pretrained phonation representations, degrading neutral-speech verification performance and hindering the large dataset used for the original training.
\begin{table}[h]
\centering
\caption{The comparison of EER scores on the testing dataset across trials of normal vs whispered speech (Norm vs Whsp), normal vs normal speech (Norm vs Norm), whispered vs whispered speech (Whsp vs Whsp), both whispered and normal vs whispered and normal speech (All vs All). The table lists mean and standard deviation from 10 independent runs. The best scores are in bold.}
\vspace{0.5em}
\begin{tabular}{|l|c|c|c|c|}
\hline
 \diagbox[width=3.2cm,height=0.7cm]{\textbf{Model}}{\textbf{Trial}} &  \textbf{Norm vs Whsp} & \textbf{Norm vs Norm} &  \textbf{Whsp vs Whsp} & \textbf{All vs All}\\
\hline

\begin{tabular}{@{}l@{}}ECAPA-TDNN \\ + post-processing 
\end{tabular}
& \begin{tabular}{@{}c@{}} 11.59\%
\\$\pm$  1.04\% \end{tabular}   
& \begin{tabular}{@{}c@{}} 0.64\% \\$\pm$ 0.17\% \end{tabular} 
& \begin{tabular}{@{}c@{}}2.46\% \\$\pm$  0.24\% \end{tabular}  
& \begin{tabular}{@{}c@{}}14.20\% \\$\pm$ 1.00\% \end{tabular}	\rule{0pt}{5ex}\\

\begin{tabular}{@{}l@{}}ReDimNet-B6 \\unfrozen last 2 blocks 
\end{tabular}
& \begin{tabular}{@{}c@{}}5.98\% \\$\pm$  0.88\% \end{tabular}   
& \begin{tabular}{@{}c@{}} 0.27\% \\$\pm$ 0.18\% \end{tabular} 
& \begin{tabular}{@{}c@{}}2.55\% \\$\pm$  0.33\% \end{tabular}  
& \begin{tabular}{@{}c@{}}9.19\% \\$\pm$ 0.96\% \end{tabular}	\rule{0pt}{5ex}\\

\begin{tabular}{@{}l@{}}ReDimNet-B6 \\ fine-tune 
\end{tabular}
& \begin{tabular}{@{}c@{}}17.00\% \\$\pm$  1.33\% \end{tabular}   
& \begin{tabular}{@{}c@{}} 5.32\% \\$\pm$ 1.06\% \end{tabular} 
& \begin{tabular}{@{}c@{}}8.53\% \\$\pm$  1.57\% \end{tabular}  
& \begin{tabular}{@{}c@{}}17.85\% \\$\pm$  1.55\% \end{tabular}	\rule{0pt}{5ex}\\

\textbf{Ours} 
& \begin{tabular}{@{}c@{}} \textbf{5.27\%} \\$\pm$ 0.82 \% \end{tabular}   
& \begin{tabular}{@{}c@{}}\textbf{0.28\%} \\$\pm$ 0.20 \% \end{tabular} 
& \begin{tabular}{@{}c@{}} \textbf{1.88\%} \\$\pm$ 0.28 \% \end{tabular}  
& \begin{tabular}{@{}c@{}}\textbf{8.40\%} \\$\pm$ 0.90 \% \end{tabular}	\rule{0pt}{5ex}\\

\hline 
\end{tabular}
\label{tab:results-ablation}
\end{table}

The primary experiment sought to determine if incorporating the speaker classification head and encoder-decoder block was beneficial during training or redundant. The findings indicate that excluding this module during fine-tuning leads to the model losing valuable speaker verification structures, resulting in a substantial increase in EER compared to the baseline result of ReDimNet-B6, with figures rising from 7.76\% to 17.85\% in the All vs All trial.

\subsection{Noise Robustness}
Additionally, we evaluate how the best baseline models ECAPA-TDNN, ECAPA2, ReDimNet-B2, and ReDimNet-B6 are affected by noise. We combine MUSAN noises with normal and whispered speech so that the peak signal to noise ratio (PSNR) is comparable for both conditions, meaning the noise added to normal speech is louder than the noise added to whispered speech.

For normal speech we choose \(SNR = \{\SI{5}{dB}, \SI{15}{dB} \}\) encompassing both low and high noise. Then, SNR for whispered speech is scaled so that PSNR for normal and whispered speech are approximately the same. The measured PSNRs are \(\mathrm{PSNR}_{normal}(\SI{5}{dB}) = 21.30\),  \(\mathrm{PSNR}_{normal}(\SI{15}{dB}) = 37.85\), \(\mathrm{PSNR}_{whisper}(\SI{5}{dB}) = 22.87\),  \(\mathrm{PSNR}_{whisper}(\SI{15}{dB}) = 37.84\).

Tab.~\ref{tab:noise-high} lists results for \(\mathrm{PSNR}(\SI{15}{dB}) \approx 38\). For all models adding noise acted detrimentally, with ReDimNet-B6 receiving the biggest drop in performance (All vs All trial) of 16.78 percentage points. The second biggest decline occured in ECAPA2 with performance being worse 13.97 percentage points than for clean samples. ECAPA-TDNN and ReDimNet-B2 dropped by 13.43 and 13.77 percentage points accordingly. 

Tab.~\ref{tab:noise-low} presents values achieved for \(\mathrm{PSNR}(\SI{5}{dB}) \approx 22\), so now the signal is more noisy. ECAPA2 and ReDimNet-B6 were not susceptible to the change in noise levels; the performance has not dropped. For ECAPA-TDNN ERR has declined by further 5.45 percentage points, totaling in 18.88 percentage points difference compared to EER with clean speech. Interestingly, ReDimNet-B2 is slightly better than for higher PSNR, improving its performance by 0.06 percentage points.

To gain more insights, we calculated relative change compared to EERs for clean speech in Norm vs Norm and Whsp vs Whsp trials for both PSNRs. The relative change is defined as:
\begin{equation}
\label{eq:relchan}
    \Delta(EER)= \frac{|EER_{\{low \ \mathrm{PSNR}, high \ \mathrm{PSNR}\}} - EER_{clean}|}{EER_{clean}} \cdot 100\%
\end{equation}

The results are listed in Tab.~\ref{tab:noise-changes}.
The relative change compared to EERs on clean speech is mostly stronger for whispered speech (expect for ECAPA-TDNN with \(PSNR\approx 22\)). ECAPA2 showed almost the same level of degradation for normal and whispered speech but for other models the difference is noticable. We can conclude that generally adding noise of the same relative power to speech, degrades performance of speaker verification more strongly with whispered speech than with clean speech.

\begin{table}[h]
\centering
\caption{The comparison of EER scores on the testing dataset with high SNR (\(\mathrm{PSNR}(\SI{15}{dB}) \approx 38\)) of baselines across trials of normal vs whispered speech (Norm vs Whsp), normal vs normal speech (Norm vs Norm), whispered vs whispered speech (Whsp vs Whsp), both whispered and normal vs whispered and normal speech (All vs All). The table lists mean and standard deviation from 10 independent runs.}
\vspace{0.5em}
\begin{tabular}{|l|c|c|c|c|}
\hline
 \diagbox[width=2.2cm,height=0.7cm]{\textbf{Model}}{\textbf{Trial}} &  \textbf{Norm vs Whsp} & \textbf{Norm vs Norm} &  \textbf{Whsp vs Whsp} & \textbf{All vs All}\\
\hline

ECAPA-TDNN 
& \begin{tabular}{@{}c@{}}27.53\% \\$\pm$ 1.35 \% \end{tabular}   
& \begin{tabular}{@{}c@{}}2.40\% \\$\pm$ 0.53 \% \end{tabular} 
& \begin{tabular}{@{}c@{}}24.18\% \\$\pm$ 2.13 \% \end{tabular}  
& \begin{tabular}{@{}c@{}}27.15\% \\$\pm$ 1.02 \% \end{tabular}	\rule{0pt}{5ex}\\

ECAPA2 
& \begin{tabular}{@{}c@{}}24.36\% \\$\pm$ 0.91 \% \end{tabular}   
& \begin{tabular}{@{}c@{}}1.70 \% \\$\pm$ 0.56 \% \end{tabular} 
& \begin{tabular}{@{}c@{}}21.81\% \\$\pm$ 1.98 \% \end{tabular}  
& \begin{tabular}{@{}c@{}}24.92\% \\$\pm$ 1.23 \% \end{tabular}	\rule{0pt}{5ex}\\

ReDimNet-B2 
& \begin{tabular}{@{}c@{}}26.51\% \\$\pm$ 1.13 \% \end{tabular}   
& \begin{tabular}{@{}c@{}}1.25\% \\$\pm$ 0.59 \% \end{tabular} 
& \begin{tabular}{@{}c@{}}24.08\% \\$\pm$ 1.76 \% \end{tabular}  
& \begin{tabular}{@{}c@{}}26.52\% \\$\pm$ 0.68\% \end{tabular}	\rule{0pt}{5ex}\\

ReDimNet-B6 
& \begin{tabular}{@{}c@{}}27.52\% \\$\pm$ 0.94 \% \end{tabular}   
& \begin{tabular}{@{}c@{}} 0.85\% \\$\pm$ 0.39 \% \end{tabular} 
& \begin{tabular}{@{}c@{}}25.60\% \\$\pm$ 1.66 \% \end{tabular}  
& \begin{tabular}{@{}c@{}} 24.54\% \\$\pm$ 0.97 \% \end{tabular}	\rule{0pt}{5ex}\\

\hline 
\end{tabular}
\label{tab:noise-high}
\end{table}

\begin{table}[h]
\centering
\caption{The comparison of EER scores on the testing dataset with low SNR (\(\mathrm{PSNR}(\SI{5}{dB}) \approx 22\)) of baselines across trials of normal vs whispered speech (Norm vs Whsp), normal vs normal speech (Norm vs Norm), whispered vs whispered speech (Whsp vs Whsp), both whispered and normal vs whispered and normal speech (All vs All). The table lists mean and standard deviation from 10 independent runs.}
\vspace{0.5em}
\begin{tabular}{|l|c|c|c|c|}
\hline
 \diagbox[width=2.2cm,height=0.7cm]{\textbf{Model}}{\textbf{Trial}} &  \textbf{Norm vs Whsp} & \textbf{Norm vs Norm} &  \textbf{Whsp vs Whsp} & \textbf{All vs All}\\
\hline

ECAPA-TDNN 
& \begin{tabular}{@{}c@{}}35.63\% \\$\pm$ 0.91 \% \end{tabular}   
& \begin{tabular}{@{}c@{}}6.14\% \\$\pm$ 0.62 \% \end{tabular} 
& \begin{tabular}{@{}c@{}}34.85\% \\$\pm$ 1.56 \% \end{tabular}  
& \begin{tabular}{@{}c@{}}32.60\% \\$\pm$ 0.85 \% \end{tabular}	\rule{0pt}{5ex}\\

ECAPA2 
& \begin{tabular}{@{}c@{}}24.36\% \\$\pm$ 0.91 \% \end{tabular}   
& \begin{tabular}{@{}c@{}}1.70 \% \\$\pm$ 0.56 \% \end{tabular} 
& \begin{tabular}{@{}c@{}}21.81\% \\$\pm$ 1.98 \% \end{tabular}  
& \begin{tabular}{@{}c@{}}24.92\% \\$\pm$ 1.23 \% \end{tabular}	\rule{0pt}{5ex}\\

ReDimNet-B2 
& \begin{tabular}{@{}c@{}}26.30\% \\$\pm$ 1.22 \% \end{tabular}   
& \begin{tabular}{@{}c@{}}1.30 \% \\$\pm$ 0.59 \% \end{tabular} 
& \begin{tabular}{@{}c@{}}23.60\% \\$\pm$ 1.87 \% \end{tabular}  
& \begin{tabular}{@{}c@{}}26.46\% \\$\pm$ 0.76 \% \end{tabular}	\rule{0pt}{5ex}\\

ReDimNet-B6 
& \begin{tabular}{@{}c@{}}27.52\% \\$\pm$ 0.94 \% \end{tabular}   
& \begin{tabular}{@{}c@{}} 0.85\% \\$\pm$ 0.39 \% \end{tabular} 
& \begin{tabular}{@{}c@{}}25.60\% \\$\pm$ 1.66 \% \end{tabular}  
& \begin{tabular}{@{}c@{}} 24.54\% \\$\pm$ 0.97 \% \end{tabular}	\rule{0pt}{5ex}\\

\hline 
\end{tabular}
\label{tab:noise-low}
\end{table}

\begin{table}[h]
\centering
\caption{Relative change of EER (defined as in Equation \ref{eq:relchan}) in noisy conditions compared to results achieved with clean speech across normal vs normal (n vs n) and whispered vs whispered (w vs w) trials.}
\vspace{0.5em}
\vspace{1em}

\begin{tabular}{|l|c|c|c|c|}
\hline
 \diagbox[width=2.2cm,height=0.7cm]{\textbf{Model}}{\textbf{Trial}} &  
 \begin{tabular}{@{}c@{}}\textbf{N vs N} \\  \(\mathrm{PSNR} \approx 38\)\end{tabular} 
 &  \begin{tabular}{@{}c@{}}\textbf{W vs W} \\  \(\mathrm{PSNR} \approx 38\)\end{tabular}
 &  \begin{tabular}{@{}c@{}}\textbf{N vs N} \\ \(\mathrm{PSNR} \approx 22\)\end{tabular} 
 &  \begin{tabular}{@{}c@{}}\textbf{W vs W} \\  \(\mathrm{PSNR} \approx 22\)\end{tabular} \\
\hline

ECAPA-TDNN 
& 3.15
&  7.73
& 13.97
& 11.58\rule{0pt}{5ex}\\

ECAPA2 
& 7.01   
& 7.79
& 7.01
& 7.79	\rule{0pt}{5ex}\\

ReDimNet-B2 
& 4.43
& 9.94 
& 4.65
& 9.72 \rule{0pt}{5ex}\\

ReDimNet-B6 
& 6.08
& 10.13
& 6.08 
& 10.13	\rule{0pt}{5ex}\\

\hline 
\end{tabular}
\label{tab:noise-changes}
\end{table}


\section{Conclusion}
In this paper, we showed that adding an encoder-decoder-like architecture on top of a speaker verification model fine-tuned and trained with a speaker classification head with cosine similarity and triplet loss improves speaker verification with whispered speech. Specifically, using ReDimNet-B6 as the encoder  enhances the EER by a relative 22.26\% compared to the baseline (baseline 6.77\% vs ours 5.27\%) in Normal vs Whispered speech trials, achieving AUC of 98.16\%. For Whispered vs Whispered tests our model achieves EER of 1.88\% with AUC equal to 99.73\%, showing 15\% relative improvement over the previous best ReDimNet-B2. We provided a comparative evaluation of current state-of-the-art speaker verification systems under whispered speech conditions. We discovered that even contemporary solutions must adapt to maintain their advanced performance, particularly with whispered speech. For instance, ReDimNet-B6's effectiveness in verifying normal vs normal speech compared to normal vs whispered speech decreased by relatively 55.42\%. We also analyzed how noise affects the speaker verification models on normal and whispered speech, obtaining that generally the same relative level of noise has a more degrading influence on whispered speech than on normal speech.

Despite the promising improvements achieved by incorporating an encoder decoder architecture into the speaker verification pipeline, several limitations remain. First, our experiments were conducted on one dataset with whispered speech, which may not capture the full diversity of real-world whispering styles, recording environments, and speaker demographics. In fact, there is a limited number of datasets with English whispered speech which hinders obtaining a general robust model. As a next step, the model may be trained using both major whispered speech datasets, CHAINS and wTIMIT. Secondly, our framework requires fine-tuning a complex speaker verification model, which could be resource-intensive in terms of computation.

Future work should explore more data-efficient or lightweight architectures that preserve performance while reducing computational cost. Further experiments could explore multilingual and cross-lingual whispered datasets to help determine the model’s adaptability across linguistic contexts. Moreover, due to scarce databases with whispered speech, collecting a larger dataset may be considered for future work to fundamentally improve the performance of speaker verification systems with whispered speech, including samples collected in real-world scenarios. Finally, adding synthetically generated data might help with generalization to new speakers and environmental conditions, which influence speaker verification with clean and whispered speech unfavorably as shown in experiments with added noise.

\begin{credits}
\subsubsection{\ackname} This work has been partially funded by Department of Artificial Intelligence, Wrocław University of Science. Created using resources provided by Wroclaw Centre for Networking and Supercomputing
(http://wcss.pl), grant no. 1754073716.

\end{credits}

%
%
%
\bibliographystyle{splncs04}
\bibliography{samplepaper}

\end{document}